\documentclass[superscriptaddress,showpacs,twocolumn,prl,floatfix]{revtex4-1}
\usepackage[english]{babel}
\usepackage{units}
\usepackage{mathrsfs}
\usepackage{amsmath}
\usepackage{amssymb}
\usepackage{graphicx}
\usepackage{esint}
\usepackage{ulem}
\usepackage{mathrsfs}
\usepackage{dcolumn}
\usepackage{bm}
\usepackage{color}
\def\be{\begin{equation}}
\def\ee{\end{equation}}
\def\ber{\begin{eqnarray}}
\def\eer{\end{eqnarray}}

\def\ev{{\bf e}}

\def\jv{{\bf j}}
\def\kv{{\bf k}}
\def\vv{{\bf v}}
\def\Mv{{\bf M}}
\def\mv{{\bf m}}

\def\pv{{\bf p}}

\def\rv{{\bf r}}

\def\vv{{\bf v}}

\def\nablabold{\mbox{\boldmath $\nabla$}}

\allowdisplaybreaks
\begin{document}

\title{Nonlocal Drag of Magnons in a Ferromagnetic Bilayer}

\author{Tianyu Liu}

\affiliation{Optical Science and Technology Center and Department of Physics and
Astronomy, University of Iowa, Iowa City, Iowa 52242, USA}

\author{G. Vignale}
\affiliation{Department of Physics and Astronomy, University of Missouri, Columbia,
Missouri 65211, USA}

\author{M. E. Flatt\'{e}}
\affiliation{Optical Science and Technology Center and Department of Physics and
Astronomy, University of Iowa, Iowa City, Iowa 52242, USA}

\begin{abstract}
Quantized spin waves, or magnons, in a magnetic insulator are assumed to interact weakly with the surroundings, and to flow with little dissipation or drag, producing exceptionally long diffusion lengths and relaxation times. In analogy to Coulomb drag in bilayer two dimensional electron gases, in which the contribution of the Coulomb interaction to the electric resistivity is studied by measuring the interlayer resistivity (transresistivity), we predict a nonlocal drag of magnons in a ferromagnetic bilayer structure based on semiclassical Boltzmann equations. Nonlocal magnon drag depends on magnetic dipolar interactions between the layers and manifests  in the magnon current transresistivity and the magnon thermal transresistivity, whereby a magnon current in one layer induces a chemical potential gradient and/or a temperature gradient in the other layer. The largest drag effect occurs when the magnon current flows parallel to the magnetization, however for oblique magnon currents a large transverse current of magnons emerges. We examine the effect for practical parameters, and find that the predicted induced temperature gradient is readily observable.\end{abstract}
\maketitle
\newpage

In a conducting bilayer system a steady current in an active layer
produces a charge accumulation in a passive layer (in which no current flows) through interlayer Coulomb
interaction. This effect, known as Coulomb drag, \cite{gramila_mutual_1991,jauho}  is caused by interlayer interactions and is essentially independent of the scattering mechanisms within the layer.
We propose here an analogous spin-wave (magnon) drag in an insulating ferromagnetic
bilayer system: the effect is caused by the magnetic dipolar interaction \textit{between the layers}  and is essentially independent of  intralayer magnon-phonon and magnon-magnon interactions. The term magnon drag was first used to describe the increase of thermopower in ferromagnetic metals at low temperature due to \textit{local} magnon-electron interactions~\cite{bailyn_maximum_1962,blatt_magnon-drag_1967,grannemann_magnon-drag_1976,kim_magnon_2008}. However, since magnons and phonons share many common features, such as their bosonic character, separating the magnon contribution from the phonon one, especially in weak magnetic fields, remains challenging. Recently, Costache \textit{et al} directly observed a local magnon drag in a thermopile formed by parallel ferromagnetic metal wires~\cite{costache_magnon-drag_2011}.
Ref.~\cite{zhang_magnon_2012} predicted that, in a layered structure, charge current flowing in one layer could induce, through spin transfer torque and local magnon drag, an electric field in a second layer. This effect was confirmed by recent experiments~\cite{wu_observation_2016,li_observation_2016}.

In contrast to Refs. [7] and [8], we consider here a \textit{nonlocal} magnon drag caused by long-range dipolar interactions between spatially separated layers.  From the Boltzmann equation we show that  a steady magnon current or magnon heat current running in the active layer induces gradients of temperature and/or magnon chemical potential in the passive layer. We emphasize that we consider incoherent distributions of magnons in each layer.   This differs from Ref.~\cite{dutta_compact_2015}, which describes a coherent ``cross-talk" effect between magnons flowing in two wires coupled by the dipolar interaction, similar to the effects of parasitic capacitance in charge-based electronics.  Far from being a parasitic effect,  the magnon transresistivity described here opens a pathway to study the contribution of dipolar interactions to magnon transport, which is important in current experiments and applications in spin-wave spintronics~\cite{serga_yig_2010,lenk_building_2011,Demokritov,Kajiwara,uchida_spin_2010,vogel_optically_2015}, in which the excited spin waves have relatively long wavelength and the dipolar interaction easily overwhelms the exchange interaction.   Furthermore, the  anisotropy of the interlayer dipolar interaction allows the induced effects on magnon transport to have components transverse to the current -- a novel effect that we name the ``nonlocal magnon drag Hall effect".

\begin{figure}[h]
    \includegraphics[width=0.8\linewidth]{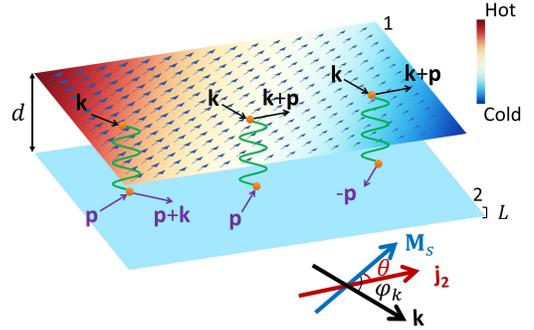} \caption{Schematic illustration of nonlocal magnon drag. The two layers have the same thickness, $L$, and are separated by a distance $d$. Layer 1 is the passive layer in which the induced temperature gradient is shown as a change of color, and the induced gradient of chemical potential is indicated by the change of arrow lengths. The magnetizations of the two layers are assumed to be in the
plane of the film and parallel to each other. The Feynman diagrams show the interlayer dipolar interactions.  }

\label{FigDrag}
\end{figure}

A schematic picture of the nonlocal magnon drag geometry is shown in Fig.~\ref{FigDrag} with two parallel layers with parallel in-plane magnetization.
The two layers are well separated and thermally
isolated, so they only interact via the magnetic dipolar interaction. In analogy to Coulomb drag, a steady magnon current flows
in the active layer (layer 2), and the passive layer (layer 1) begins
in equilibrium (i.e., there is neither magnon current nor heat current).
The magnon current density is defined by the magnon distribution
function. Before switching on the steady current both layers are in thermal
equilibrium at the same temperature $T$. A steady magnon current in layer
2 brings its own distribution out of equilibrium, and drives layer
1 to a new quasi-equilibrium state through the interlayer dipolar interaction.
The change in the distribution function in layer 1 reflects the formation of a chemical potential gradient, $\nablabold \mu_1$, and a temperature gradient, $\nablabold T_1$. The induced gradients in layer 1 are connected to the driving magnon current $\jv_2$, and magnon heat current $\jv_{Q2}$, by the $2\times 2$ transresistivity matrix ($\underline{C}^{12}$, where underline indicates the symbol denotes a matrix with scalar matrix elements):
\ber
\begin{pmatrix}\nablabold\mu_{1}\\
\frac{\nablabold T_{1}}{T_{1}}
\end{pmatrix}&=&-\underline{C}^{12}\cdot\begin{pmatrix}j_{2}\\
j_{Q2}
\end{pmatrix}\left[\frac{7}{16} (\hat\jv_2\cdot\hat \Mv_s) \hat \Mv_s\right.\nonumber\\
&& \left.-\frac{3}{16}(\hat\jv_2\times\hat\Mv_s)\times\hat \Mv_s\right]\,,\label{rou12}
\eer
 where $\hat{\jv}_2$ and $\hat{\Mv}_s$ are the unit vectors of the driving currents ($\jv_2$ and $\jv_{Q2}$, which are in the same direction) and the saturation magnetization ($\Mv_s$), respectively. Equation (\ref{rou12}) shows that the induced field is maximal when the driving current is parallel to $\Mv_s$ and is minimal when the two are perpendicular. It is only in these two cases that the induced fields are parallel to the driving current.  For all other directions, the induced field in the passive layer will have a component perpendicular to the direction of the driving current.

The transresistivity matrix $\underline{C}^{12}$ and the angular-dependence are determined by solving the Boltzmann equations for the two layers. For layer 1,
\begin{equation}
\bigg(-\frac{\partial n_{1}^{0}}{\partial\epsilon_{\kv}}\bigg)\vv_{\kv}\cdot\left(\nablabold\mu_{1}+\frac{\tilde{\epsilon}_{1\kv}}{T}\nablabold T_{1}\right)=\bigg(\frac{\partial n_{1}(\rv,\kv,t)}{\partial t}\bigg)_{12}\,,\label{BE}
\end{equation}
where $\tilde{\epsilon}_{1\kv} \equiv \epsilon_{1\kv} -\mu_1$, $\epsilon_{1\kv}$ magnon dispersion in layer 1.
The interlayer collision term on the right hand side of Eq. (\ref{BE})
is constructed by applying Fermi's golden rule to the dipolar interaction
\begin{eqnarray}
\mathscr{\mathscr{H}}_{dip} & = & \frac{\mu_{0}}{8\pi}\int_{d-\frac{L}{2}}^{d+\frac{L}{2}}d\xi\int d\rv\int_{-\frac{L}{2}}^{\frac{L}{2}}d\xi'\int d\rv'\nonumber\\
&&\frac{[\nabla\cdot\mv(\rv,\xi)][\nabla'\cdot\mv(\rv',\xi')]}{\sqrt{|\rv-\rv'|^{2}+(\xi-\xi')^{2}}}\,,
\end{eqnarray}
where $L$ is the thickness of each layer, $\rv$ is the position vector within the film plane, $\xi$ is the
coordinate perpendicular to the film, and $\mv$ is the small deviation of the magnetization due to the existence of the magnons.

The explicit form of the collision term  (see Supplemental Material~\cite{suppl})\nocite{holstein_field_1940,kalinikos_theory_1986} is
\begin{eqnarray}
&&\bigg(\frac{\partial n_{1}}{\partial t}\bigg)_{12}\nonumber\\
 &=&-\sum_{\pv}\frac{2\pi}{\hbar}|W(\kv)|^{2}(n_{1\kv}n_{2\pv}\bar{n}_{2,\pv+\kv}-\bar{n}_{1\kv}\bar{n}_{2\pv}n_{2,\pv+\kv})\nonumber\\
 &&\times\delta(\epsilon_{1\kv}+\epsilon_{2\pv}-\epsilon_{2,\pv+\kv})\nonumber \\
 &&-\sum_{\pv}\frac{2\pi}{\hbar}|W(\pv)|^{2}\left[(n_{1\kv}\bar{n}_{1,\kv+\pv}n_{2\pv}-\bar{n}_{1\kv}n_{1,\kv+\pv}\bar{n}_{2,\pv})\right.\nonumber\\
 &&\times\delta(\epsilon_{1\kv}+\epsilon_{2\pv}-\epsilon_{1,\kv+\pv})\nonumber \\
 && +(n_{1\kv}\bar{n}_{1,\kv+\pv}\bar{n}_{2,-\pv}-\bar{n}_{1\kv}n_{1,\kv+\pv}n_{2,-\pv})\nonumber\\
 &&\left.\times\delta(\epsilon_{1\kv}-\epsilon_{2\pv}-\epsilon_{1,\kv+\pv})\right]\,,\label{eq:coll}
\end{eqnarray}
with $\bar{n}_{i,\kv}\equiv1+n_{i,\kv}$ $(i=1,2)$, and
\begin{eqnarray}
W(\kv)&=&\frac{\sqrt{2}}{8}\mu_{0}\left(\frac{g\mu_{B}}{L}\right)^{\frac{3}{2}}\frac{\sqrt{M_{s}}}{k}e^{-k(d-L)}(1-e^{-kL})^{2}\nonumber\\
&&\times\cos\varphi_{\kv}(1+\sin\varphi_{\kv})\,,\label{eq:W}
\end{eqnarray}
where $g$ is the Land\'{e} factor, $\mu_B$ is the Bohr magneton, $d$ is the distance between the two layers,  and $\varphi_{\kv}$ is the angle between $\kv$ and
$\Mv_{s}$. The tunneling of magnons between the two layers
has been suppressed by requiring that the magnons in the two layers
have different dispersions. For example, the two layers
can be subject to different Zeeman fields or they can be made of materials
of different exchange stiffnesses. Here we assume the former condition so that the magnons in the two layers have different energy gaps, $\epsilon_{01}$ and $\epsilon_{02}$. The Boltzmann equation for layer 2 is obtained by interchanging subscripts 1 and 2 and $\kv$ and $\pv$ in Eqs. (\ref{BE}) and~(\ref{eq:coll}).

The transresistivities depend on the momentum ($\kv$) and thermal momentum ($\kv (\epsilon_\kv-\mu)$) transfer rates, obtained by multiplying the collision terms by $\kv$ and \hbox{$\kv (\epsilon_\kv-\mu)$} respectively and performing the summation over $\kv$. By linearizing the collision terms about the regular and thermal drift momenta, driving currents are introduced into the transfer rates, which, in a steady state, must be balanced by opposite rates of change generated by the  fields in each layer coming from the left hand side of the Boltzmann equations. From these balance conditions we obtain the equations that connect  \hbox{($\jv_{1}$, $\jv_{Q1}$, $\jv_{2}$, $\jv_{Q2}$)$^T$} to  \hbox{($\nablabold\mu_{1}$, $\nablabold T_{1}/T_1$, $\nablabold\mu_{2}$, $\nablabold T_{2}/T_2$)$^T$,} as explained in detail in the Supplemental Material~\cite{suppl}.

The momentum transfer rate due to the interlayer collision term, obtained by multiplying Eq.~(\ref{eq:coll}) by $\kv$ and summing over $\kv$, depends on the summation
\ber\label{eq:sum}
&&\sum_{k}\int d\varphi_{\kv}\hat{\kv}(\hat{\kv}\cdot\jv_{2})f(k)\cos^2\varphi_{\kv}(1+\sin\varphi_{\kv})^{2}\nonumber\\
&=&\sum_k f(k) j_2\left[\frac{7}{16} (\hat\jv_2\cdot\hat \Mv_s) \hat \Mv_s-\frac{3}{16}(\hat\jv_2\times\hat\Mv_s)\times\hat \Mv_s\right] \nonumber\\
&=&\sum_k f(k) j_2\{[(3+4\cos^2\theta)/16]\hat{\ev}_{\parallel}-(1/8)\sin 2\theta \hat{\ev}_{\perp}\}\label{eq:1}
\eer
where the $\cos^2\varphi_{\kv}(1+\sin\varphi_{\kv})^{2}$ in the integrand comes from the square of the interlayer dipolar interaction, Eq.~(\ref{eq:W}), $f(k)$ denotes a function that depends only on the magnitude of $\kv$, $\theta$ is the angle between $\hat{\jv}_2$ and $\hat{\Mv}_s$ as shown in Fig. \ref{FigDrag},  and $\hat{\ev}_{\parallel}$ ($\hat{\ev}_{\perp}$) is the unit vector parallel (perpendicular) to $\hat \jv_2$. A similar summation appears in the calculation of the ``thermal momentum" transfer rate. Equation (\ref{eq:sum}) shows that the $\theta$ dependence of the transresistivities results from the anisotropy of the interlayer dipolar interactions.

\textit{Temperature dependence of the transresistivities.}---There are two general approaches to the study of thermodynamic properties of quasi-particles:
at fixed chemical potential (grand canonical ensemble) or at fixed particle number (canonical ensemble).  Yttrium iron garnet (YIG) is a proper material for these studies. Due to its short thermalization time ($<\unit[100]{ns}$) and relative long spin-lattice relaxation time ($>\unit[1]{\mu s}$), magnons in YIG can maintain a quasi-equilibrium state with a non-zero chemical potential via energy transfer from microwave fields through parametric pumping~~\cite{demidov_thermalization_2007,dzyapko_direct_2007}. Noting that the chemical potential increases with increased  pumping power at a given temperature (i.e. room temperature, as in Ref. \cite{dzyapko_direct_2007}), one can realize a fixed chemical potential $\mu_{2}$ in the active layer by adjusting the pumping power for different temperatures. On the other hand, a fixed number of pumped magnons can be achieved by fixing the frequency and the power of the microwave field. In both cases, the transresistivities depend on the temperature $T$ of the active layer.

\textit{Fixed chemical potential.}---Suppose a steady magnon current flows in layer 2 along $\Mv_{s}$ ($\theta=0$ for a maximal effect), and the two layers are initially at the
same temperature ($T$). The transresistivities
have been defined in Eq. (\ref{rou12}). In the high temperature limit such that $(\epsilon_{0i}-\mu_{i})/k_{B}\ll T\ll T_{c}$ ($T_{c}$ is the Curie temperature and is $550$ K for YIG), we consider only the momentum transfer rate due to the interlayer collision and the leading order terms in temperature.  A quadratic dispersion ($\epsilon_{ik}=Dk^2+\epsilon_{0i}$, with $D$ the exchange stiffness) has been assumed to simplify the calculations. The resulting transresistivity matrix is
\be
\underline{C}^{12}=\left[\frac{9\hbar\mu_{0}^2}{2}\bigg(\frac{g\mu_{B}}{L}\bigg)^{3}\frac{M_{S}}{\epsilon_{02}-\mu_{2}}\right]\underline{\tilde C}^{12}\,,
\ee
and the temperature dependence of its components (in the high-temperature regime)  is
\begin{eqnarray}
\tilde{C}_{\mu\mu}^{12} & \propto & {D^{2}\pi^{2}(18\zeta[3])^{2}}[\Theta(T)]^{-1}\,,\label{eq:Cmumu2}\\
\tilde{C}_{\mu T}^{12} & = & \tilde{C}_{T\mu}^{12}\propto{-D^{2}T^{-1}\pi^{2}18\zeta[3]}[\Theta(T)]^{-1}\,,\\
\tilde{C}_{TT}^{12} & \propto & {D^{2}T^{-2}\pi^{4}}[\Theta(T)]^{-1}\,,\label{eq:CTT2}
\end{eqnarray}
where $\zeta$ is the Riemann zeta function and \hbox{$\Theta(T)=\{\pi^{4}+54\ln(\beta\epsilon_{01})\zeta[3]\}\{\pi^{4}+54\ln[\beta(\epsilon_{02}-\mu_{2})]\zeta[3]\}$.} The full expression for $\underline{\tilde{C}}^{12}$ is in the Supplemental Material~\cite{suppl}.

\begin{figure}
\includegraphics[width=1\linewidth]{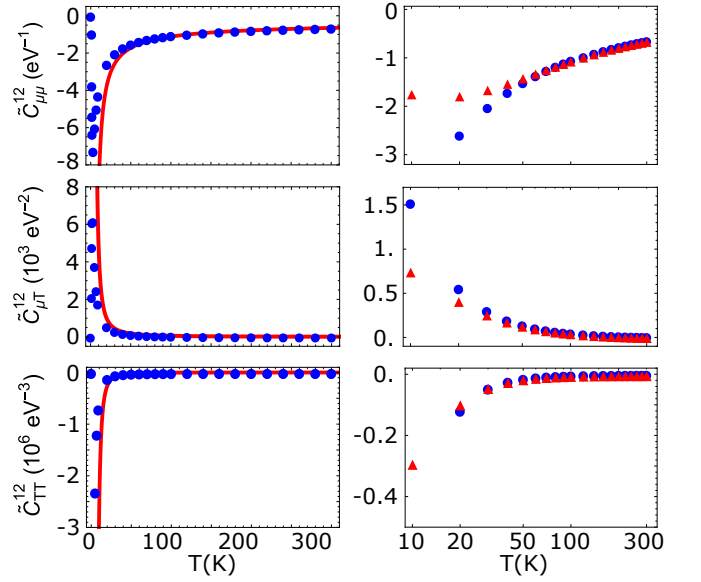}

\caption{The matrix elements of $\underline{\tilde{C}}^{12}$ as a function of $T$.
Left column: dots are the exact numerical results and solid lines are the
analytical results in the high temperature approximation (Eqs.~(\ref{eq:Cmumu2})-(\ref{eq:CTT2})) for a fixed chemical potential. Right column: blue dots are for fixed chemical potential; red triangles are for fixed number of pumped magnons. The parameters that have been used to generate these
plots are $\epsilon_{01}=0.5$ K, $\epsilon_{02}=1$ K, $\mu_{1}=0$
K, $d=6$ nm, $L=3$ nm, $k_{B}=1$,
and $D=3.01\times10^{-17}\unit{K\cdot m^{2}}$; $\mu_{2}=0.2$ K for the left column, and $\delta n_2=1.8\times10^{17}\unit{\,m^{-2}}$ for the right column.\label{fig:CT}}.
\end{figure}

The matrix elements of $\underline{\tilde{C}}^{12}$ as a function of $T$, shown in Fig. \ref{fig:CT}, are well described by the high temperature approximation for $T>10$ K (i.e. $(\epsilon_{0i}-\mu_{i})/(k_{B}T)<0.1$
with the parameters given in the caption). With decreasing  temperature there is
a sharp increase of the transresistivities since $\Theta(T)\propto (\ln T)^2$ from its definition given above. The transresistivities reach a maximum, and then begin to decrease. The decrease for lower temperatures  occurs because there are very few magnons thermally excited when
$T\ll(\epsilon_{0i}-\mu_i)/k_B$. In the extreme case, when $T=0$ K, there are no magnons at all and
the transresistivities vanish.

\textit{Fixed number of pumped magnons.}---The areal density of magnons in two dimensions can be
calculated analytically,
\begin{eqnarray}
n & = & \sum_{\kv}\frac{1}{e^{\beta(Dk^{2}+\epsilon_{0}-\mu)}-1}=\frac{-\ln\left[1-e^{\beta(\mu-\epsilon_{0})}\right]}{4\pi\beta D}\,.
\end{eqnarray}
At thermal equilibrium, the chemical potential $\mu=0$. Let $\delta n$
be the number of pumped magnons per unit area, then $\delta n=n(\mu,T)-n(0,T)$
and the chemical potential depends on $\delta n$ and $T$,
\begin{eqnarray}
\mu(\delta n,T) &=&\epsilon_{0}+ \frac{1}{\beta}\ln\left[1-\left(1-e^{-\beta\epsilon_{0}}\right)e^{-4\pi\beta D\delta n}\right].
\end{eqnarray}
$\mu$ increases as $\delta n$ increases or $T$ decreases (as the magnon gas approaches Bose-Einstein condensation (BEC)) and
is always less than $\epsilon_{0}$. In the right column of Fig. \ref{fig:CT} we compare the transresistivities calculated for a fixed
chemical potential $\mu_{2}$ (blue dots) and those for a fixed number
of pumped magnons (red triangles). The results show that the most
significant difference is in $\tilde{C}_{\mu\mu}^{12}$ in the low
temperature range. If the number of pumped magnons is fixed, lowering
the temperature will lead to the formation of a BEC below $\unit[9]{K}$ (for the parameters we are using), and
therefore, the transresistivities cannot be enhanced as much as in the fixed chemical potential case (before they begin to fall again at even lower temperatures).
The BEC is beyond the range of validity of our model,
since the magnon number in the BEC state cannot be described by the Bose-Einstein
distribution alone~\cite{rezende_theory_2009}.

We now estimate how large a $\nablabold\mu_{1}$ can be induced by $\jv_{2}$ from Eq. (\ref{rou12}).
For layer 2 at a uniform temperature, $\jv_{Q2}$ is related to $\jv_{2}$ by $\jv_{Q2}=\pi/(12n_{2}D\beta^{2})\jv_2$.
At room temperature ($n_{2}=4.7\times10^{18}$ m$^{-2}$), $j_{Q2}/j_{2}\sim0.01$
eV, and $C_{\mu T}^{12}/C_{\mu\mu}^{12}\sim-17$ eV$^{-1}$.  The contribution of the thermal current is negligible,  so that $\nablabold\mu_{1}=-(7/16)C_{\mu\mu}^{12}\jv_{2}$,
with $C_{\mu\mu}^{12}=-8.5\times10^{-42}(\unit{J\cdot s})$ for the parameters we have chosen. Suppose the magnon spin current in layer 2 is carried by pumped magnons with a density of $10^{19}\unit{\,cm^{-3}}$ and an average velocity $100\unit{\,m/s}$,
then $j_{2}=nLv\sim3\times10^{18}\unit\,{m^{-1}s^{-1}}$: the resulting $\nabla\mu_1=1.1\times10^{-23}\unit{\,J/m}$, which is difficult to observe. For example, in a recent measurement of magnon BEC \cite{dzyapko_boseeinstein_2011}, a change of chemical potential ($0.1\unit{\,mK}$) could be measured across a spot of radius $400\unit{\,nm}$. The corresponding chemical potential gradient was about $10^{-21}\unit{\,J/m}$.

Now let us turn to the induced temperature gradient in the passive
layer. Noting that $j_{Q2}/j_{2}\sim0.01$ eV and $C_{TT}^{12}/C_{T\mu}^{12}\sim-17$
eV at room temperature, we again neglect the thermal current and get $\nabla T_{1}\simeq-(7/16)C_{T\mu}^{12}j_{2}T=\unit[-0.37]{\,K/m}$, with $C_{T\mu}^{12}=4.1\times10^{-22}\unit{\,s}$. This temperature gradient is sufficiently large that it may be detected by means of the spin Seebeck effect using current experimental sensitivity \cite{uchida_spin_2010}. Suppose the magnon thermal mean free path in layer 1 is $10\unit{\,\mu m}$, the spin Hall angle in Pt is $\theta_{SH}=j_e/j_s=10 e/(g\mu_B)$, and all the magnon current is absorbed by the detector. The charge current density in the Pt contact due to the temperature gradient induced in layer 1 is about $10^6\unit{ A/m^2}$. The corresponding ISHE voltage across Pt of width $200\unit{\, \mu m}$ is $20\unit{\,\mu V}$ with a conductivity of $10^7\unit{\,(\Omega\cdot m)^{-1}}$, which is readily observable. The observed voltage in current ISHE measurements can be as small as $10\unit{\, nV}$~\cite{jin_effect_2015}, so our effect is three orders of magnitude above the sensitivity threshold.

The total effect in layer 1 is reflected in the redistribution of the magnon density,
\be
\nabla n_1=\frac 7 {16L}\sum_{\kv}\bigg(\frac{\partial n_{\kv}}{\partial\epsilon_{\kv}}\bigg)\left(C_{\mu\mu}^{12}+\frac{C_{T\mu}^{12}}{T}\right)j_2\,.
\ee
From the above analysis, we estimate the induced gradient of magnon density in layer 1 to be $2.3\times10^{26}\unit{\,m^{-4}}$. A change of the density over a distance of $1\unit{\,\mu m}$ should be detectable by micro-focused Brillouin light scattering (BLS) with sensitivity of $10^{14}\unit{\,cm^{-3}}$~\cite{Demokritov, dzyapko_boseeinstein_2011,sebastian_micro-focused_2015}, corresponding to a gradient of $10^{26}\unit{\,m^{-4}}$. Thus the predicted effect in the magnon density is more than a factor of two larger than the current experimental sensitivity.

\begin{figure}
\includegraphics[width=0.8\linewidth]{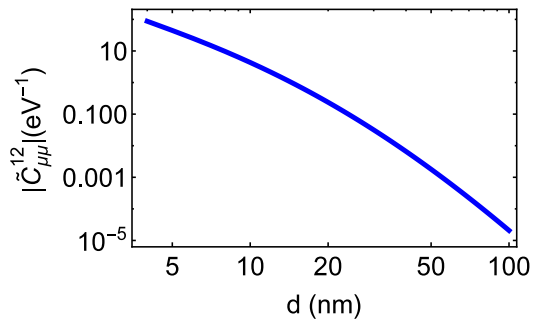}

\caption{The distance dependence of the transresistivity with parameters $\epsilon_{01}=0.5$ K, $\epsilon_{02}=1$ K, $\mu_{1}=0$
K, $\delta n_2=1.8\times10^{17}\unit{\,m^{-2}}$, $L=3$ nm, $k_{B}=1$, $D=3.01\times10^{-17}\unit{K\cdot m^{2}}$, and $T=300$ K. \label{fig:Cd}}
\end{figure}

\textit{Dependence of the transresistivities on distance between layers}--The four transresistivities (Eqs. (\ref{eq:Cmumu2}) to (\ref{eq:CTT2})) are affected by the interlayer distance via the collision term (Eqs. (\ref{eq:coll}) and (\ref{eq:W})), and share the same $d$ dependence in the high temperature limit, since they depend on the momentum and thermal momentum transfer rates induced by the regular drift momentum only (which is shown clearly in the full expression of the transresistivities in the Supplemental Material~\cite{suppl}). The dependence of $\tilde{C}_{\mu\mu}^{12}$ on $d$ is shown in Fig. \ref{fig:Cd}. For a large distance between the layers the magnon drag transresistivity is found to decrease faster than the familiar Coulomb drag transresistivity, i.e., faster than $d^{-4}$~\cite{jauho}.

To conclude, we have studied the nonlocal drag effect in a bilayer of two-dimensional
magnon gases. It behaves quite differently from ordinary Coulomb drag, due to the combined action of Bose-Einstein statistics, non-conservation of the
magnon number, and the anisotropy of the interlayer interaction.
The drag effect exhibits a strong angular-dependence. The induced fields are largest when the current flows parallel to the saturation magnetization, and have components transverse to the current for oblique flows. The transresistivities can be increased by orders of magnitude by lowering the temperature or by decreasing the interlayer distance. Although the induced chemical potential gradients are about an order of magnitude smaller than current experimental sensitivity, the induced temperature gradients we calculate are three orders of magnitude larger, and induced magnon density changes are a factor of two larger than current experimental sensitivity and thus should be readily observable.
To illustrate the drag effect quantitatively, we have considered only a quadratic magnon dispersion in this Letter. Realistic dispersion of magnons in ferromagnetic thin layers may affect our estimated quantities and deserves future consideration.

We acknowledge support of the Center for Emergent Materials, a NSF MRSEC under Award No. DMR-1420451 and an ARO MURI.

\bibliography{Ref}

\end{document}